\documentclass[aps,pra,twocolumn,showpacs,superscriptaddress,groupedaddress]{revtex4}  

\usepackage{graphicx}  
\usepackage{dcolumn}   
\usepackage{bm}        
\usepackage{amssymb}   

\usepackage{xspace}
\usepackage{amsmath}
\usepackage{abbrevs}
\usepackage[]{units}	

\usepackage{float} 
\usepackage{wrapfig} 

\newabbrev\QI{Quantum Illumination (QI)}[QI]
\newabbrev\SFWM{spontaneous four-wave mixing (SFWM)}[SFWM]
\newabbrev\NRC{National Research Council (NRC)}[NRC]
\newabbrev\QPSS{quantum photonic sensing and security (QPSS)}[QPSS]
\newabbrev\CI{Classical Illumination (CI)}[CI]
\newabbrev\SNR{signal-to-noise ratio (SNR)}[SNR]
\newabbrev\SPDC{spontaneous parametric downconversion (SPDC)}[SPDC]
\newabbrev\FWM{four-wave mixing (4WM)}[4WM]
\newabbrev\TBP{time-bandwidth product (TBP)}[TBP]
\newabbrev\APD{avalanche photodiode (APD)}[APD]

\makeatletter
\renewcommand\maybe@space@{%
  \maybe@ictrue 
  \expandafter   \@tfor
    \expandafter \reserved@a
    \expandafter :%
    \expandafter =%
                 \nospacelist
                 \do \t@st@ic
  \ifmaybe@ic 
    \space
  \fi
}
\makeatother

\begin{document}



\title{Quantum-enhanced standoff detection using correlated photon pairs}

\author{Duncan G. England} \affiliation{National Research Council of Canada, 100 Sussex Drive, Ottawa, Ontario, K1A 0R6, Canada}
\author{Bhashyam Balaji}\affiliation{Radar Sensing and Exploitation Section,
Defence R\& D Canada, Ottawa Research Centre, 3701 Carling Avenue, Ottawa, Ontario, K1A 0Z4, Canada}
\author{Benjamin J. Sussman} \email{ben.sussman@nrc.ca}\affiliation{National Research Council of Canada, 100 Sussex Drive, Ottawa, Ontario, K1A 0R6, Canada}
\affiliation{Department of Physics, University of Ottawa, 598 King Edward, Ottawa, Ontario K1N 6N5, Canada}
\date{\today}

\begin{abstract}
\noindent We investigate the use of correlated photon pair sources for the improved quantum-level detection of a target in the presence of a noise background. Photon pairs are generated by spontaneous four-wave mixing, one photon from each pair (the {\em herald}) is measured locally while the other (the {\em signal}) is sent to illuminate the target. Following diffuse reflection from the target, the signal photons are detected by a receiver and non-classical timing correlations between the signal and herald are measured in the presence of a configurable background noise source. Quantum correlations from the photon pair source can be used to provide an enhanced signal-to-noise ratio when compared to a classical light source of the same intensity.

\end{abstract}

\maketitle
\section{Introduction}


Light, and more generally electromagnetic radiation, continues to be a primary mechanism for ranging and imaging of objects at a distance.  Physical limits exist on measurements performed with classical sources emitting thermal or coherent states.  For example, Rayleigh criterion limits spatial resolution or Poisson statistics constrains the limits of shot noise processes.  Considerable effort has been taken to understand the opportunity that non-classical illumination affords for ranging, imaging, and measurement in general~\cite{HanburyBrown1956,Holland1993,Kwiat1995a,Mitchell2004,Giovannetti2004,Gatti2004,Nagata2007,Brida2010,Kacprowicz2010,LIGOSC2011,Aasi2013,Schwartz2013,Lemos2014}.  Here we investigate the use of non-classically correlated photon pairs for imaging, and demonstrate experimentally that that the signal to noise ratio can be improved beyond the classical limit by a multiple of the second-order coherence factor $g^{(2)}$.

The theoretical background for this work was developed by Lloyd in 2008~\cite{Lloyd2008}. The proposed \QI protocol used a pair of photons --- the so-called {\em signal} and {\em ancilla} photons --- that are entangled in some degree of freedom, in this case frequency. The signal photon is used to illuminate a target, while the ancilla is stored locally. The signal photon scatters from the target and, when it returns, an entanglement measurement is performed to determine whether or not entanglement remains between the signal and ancilla photons. In the presence of a large background, the number of signal photons returning to the detector can be orders of magnitude lower than the noise floor, but the joint measurement with the ancilla photons provides a means to increase the \SNR. Because the signal is strongly (and non-classially) correlated with the ancilla and the noise is not, correlation measurements can separate the two.

In the years following Lloyd's proposal, a number of theoretical papers evaluated the \QI protocol (see for example~\cite{Tan2008,Guha2009,Shapiro2009,Barzanjeh2015}). While the quantum transmitter (entangled photon source) is a well-established technology~\cite{Kwiat1995} the optimal quantum receiver (entanglement measurement) is complex and so experimental demonstrations proved more challenging. The difficulty of entanglement measurement is further increased when the signal photon is scattered from an unknown object. Accordingly, laboratory demonstrations of entanglement-based QI have so far been limited to experiments where the the signal photon remains in optical fiber to preserve coherence, and noise and loss are artificially added~\cite{Zhang2015}. Alternatively, it is possible to observe enhancement over classical schemes by simply measuring non-classical correlations rather than entanglement, resulting in far simpler apparatus, at the expense of reduced sensitivity. This approach was taken by Lopaeva {\em et al.}~\cite{Lopaeva2013} who used a \SPDC photon pair source to detect a specular reflection from a beamsplitter. In this case, the non-classical spatial correlations measured on a single-photon CCD camera were used to distinguish signal from background light in a way that provided enhanced sensitivity compared to classical light of the same intensity.

In this paper we continue to develop QI from a practical perspective by using it to detect a diffusely reflecting target. Illuminating a diffuse reflector rather than the specular reflector increases the realism of the scenario, but introduces certain challenges. In particular, collecting photons after diffuse reflection becomes increasingly difficult with distance from the target, limiting the range over which QI can be effective. As well, diffuse reflection will scramble the spatial correlations that were previously employed, so the method presented here uses, instead, only temporal correlations. Despite these challenges, a clear advantage of QI is demonstrated when compared to \CI of the same intensity over modest table-top distances. This work shows that the quantum advantage identified in reference~\cite{Lopaeva2013} is accessible using only a single detector. A simple model is developed to quantify the improvement offered by a photon pair source and  regimes in which QI could be advantageous are identified. We conclude by suggesting a route forwards for extending the useful range of QI beyond the laboratory and into a real environment.  

\section{Description of the apparatus}

\begin{figure}[h]
\center{\includegraphics[width=1.0\linewidth]{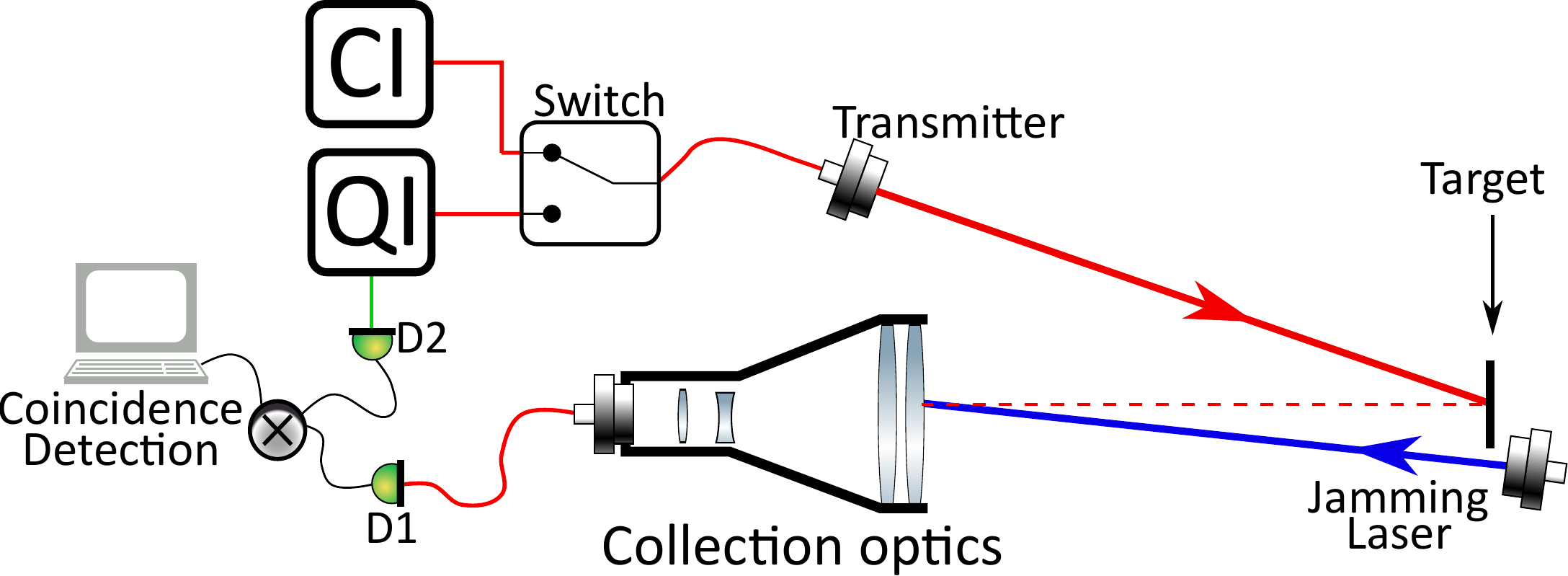}}
\caption{A schematic layout of the experiment. A source of either quantum (QI) or classical (CI) light is directed to the transmitter and illuminates a target. Light scattered from the target is collected by the collection optics and directed to detector D1. For QI, the herald beam is detected by detector D2 and coincidence measurements between the two photons are made. A second laser, the jamming laser, is used as a background light source.}
\label{fig:Overview}
\end{figure}

Our apparatus begins with a photon pair source, one photon mode (the herald) is detected locally, and the other (the signal) is scattered from a target and detected by a receiver. Background light is provided by the `jamming laser' which illuminates the receiver from behind the target. QI is performed by measuring timing statistics between the signal and herald and a comparison is made with \CI of the same intensity. An overview of the apparatus used in this demonstration is shown in figure~\ref{fig:Overview}. In the following sections, each element is described in detail.

\subsection{Target}

The target is a white card. Diffuse reflection from the target will disperse the photons, which makes collecting the scattered photons a significant challenge. The target is mounted on an re-purposed laser safety shutter so it can be moved in and out of the beam. This allows us to rapidly compare the photon flux with/without the target in place. 

\subsection{Photon pair source}

The \QI source used for these experiments is a \SFWM photon pair source based on a birefringent optical fiber~\cite{Smith2009}. Spontaneous four-wave mixing (SFWM) is a third-order ($\chi^{(3)}$) nonlinear optical process that can occur when a strong pump pulse enters a nonlinear medium. With low probability, two photons at the pump wavelength will be annihilated and a pair photons will be created, referred to as the signal and idler photons. In this demonstration, the signal photon is used to illuminate the target and the idler photon is measured locally to `herald' the generation of a signal photon. The specific wavelengths of the signal and herald photons depend upon energy conservation and phasematching in the fiber, in this case, a pump wavelength of 793\,nm was used resulting in signal and herald photon wavelengths of 671\,nm and 970\,nm respectively.

\begin{figure}[h]
\center{\includegraphics[width=0.75\linewidth]{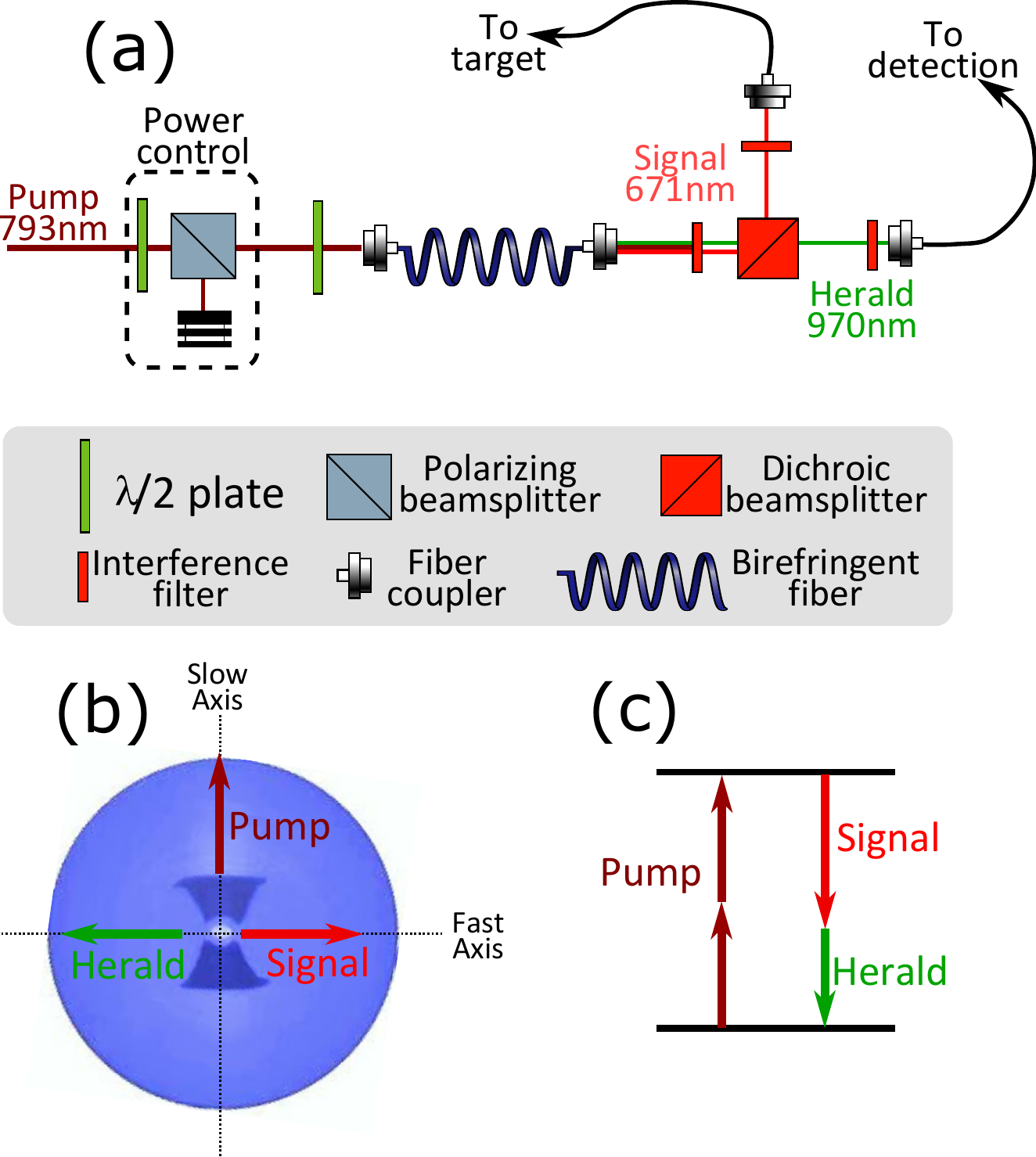}}
\caption{{\bf (a)} Schematic of the photon source. A pump laser is coupled into a 20\,cm long birefringent fiber. Photon pairs --- signal and herald --- are generated in the fiber and are spectrally isolated before being coupled into separate fibers. {\bf (b)} A microscope image of the birefringent fiber showing fast and slow axes. The pump polarization is aligned to the slow axis, and the signal/herald to the fast axis. {\bf (c)} Energy level diagram for SFWM.  }
\label{fig:setup_source}
\end{figure}

A schematic of the photon source is shown in figure~\ref{fig:setup_source}(a). The pump laser is a frequency-filtered Titanium Sapphire modelocked oscillator with central wavelength of 793\,nm, a pulse duration of $\sim 1$\,ps, a maximum pulse energy of 2.5\,nJ and a pulse repetition frequency of $R_p =80$\,MHz. A motorized $\lambda/2$ waveplate and a polarizing beamsplitter allow us to control the laser power entering the fiber, a second $\lambda/2$ waveplate is then used to align the pump polarization to the slow axis of the fiber. The birefringent fiber is the HB800 manufactured by Fibercore; it is 20\,cm long and has a birefringence of $\Delta n \simeq 3\times10^{-4}$. At the output of the fiber, a notch filter is used to remove the pump light and a dichroic mirror is used to separate the signal and idler. The signal photon passes through a 3\,nm bandpass filter centered at 671\,nm, and is then coupled into a single mode fiber and sent to the transmitter. The herald photon passes through a 10\,nm bandpass filter centered at 970\,nm before being coupled into a single mode fiber and sent to an avalanche photodiode (APD) for detection.

\begin{figure}[h]
\center{\includegraphics[width=0.95\linewidth]{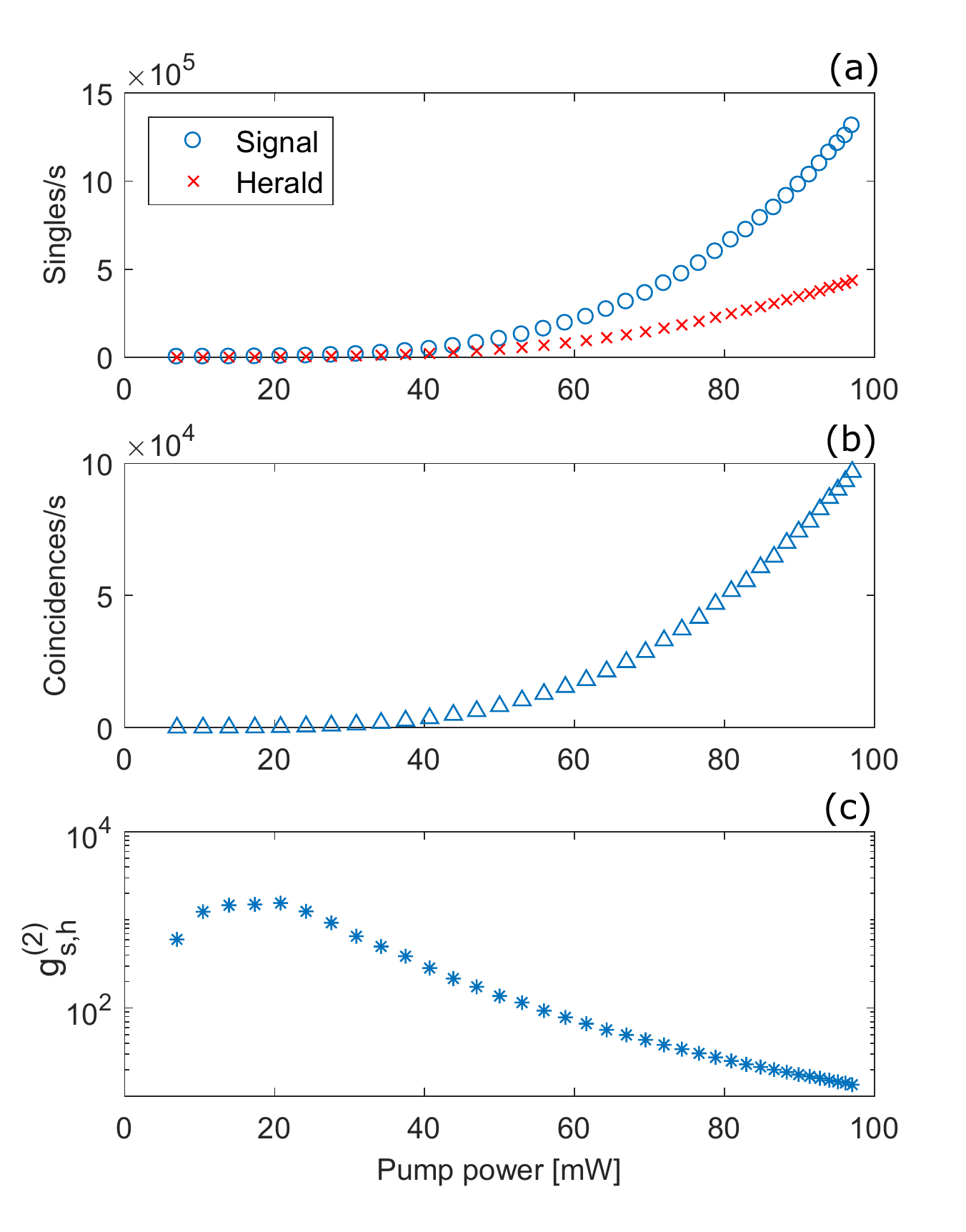}}
\caption{Photon source characterization. Signal and herald singles counts {\bf (a)}, coincidence counts {\bf (b)} and  degree of second order coherence ($g^{(2)}_{s,h}$) {\bf (c)} as a function of pump laser power.  }
\label{fig:countrates}
\end{figure}

Before using the source for a QI protocol, its performance is characterized: The signal and herald fibers are connected directly to APDs and the electrical output from each APD is then sent to a time-tagging unit for single and conincident detection measurements. The power in the pump beam is varied from 0 to 97\,mW (as measured at the fiber output) and the  number of herald and signal photons are counted, as well as the number of coincident detection events. Typical count rates are shown in figure~\ref{fig:countrates}, at maximum power $N_s\simeq1.3\times10^6$ signal photons are detected, and $N_h\sim4.4\times10^5$ herald photons. This discrepancy is largely due to the fact that the APD efficiecny at 970\,nm ($\eta_{dh} \simeq 20\%$) is lower than at 671\,nm ($\eta_{ds} \simeq 60\%$). Up to 97,000 coincident detection events are measured per second. The coincidence rate is lower than the herald rate primarily because of collection and detection inefficiencies, but also due to competing nonlinear optical processes such ({\em e.g.} Raman scattering) that create photons at the herald wavelength without a corresponding signal photon. The mean photon number, $\mu$, is the average number of signal photons generated by each laser pulse and is calculated by $\mu = N_s/(R_p\times\eta_{ds})$. The mean photon number can be varied from 0 to 0.025 by adjusting the power of the pump laser.

\subsubsection{Degree of second order coherence ($g^{(2)}_{s,h}$)} \label{sec:g2}

The strength of the correlations between signal and herald photons is given by the two-mode degree of second-order coherence $g^{(2)}_{s,h}$. When $\mu<<1$, $g^{(2)}_{s,h}$ can be measured using single photon detectors in the following way:
\begin{equation}
\label{eq:g2}
g^{(2)}_{s,h} = \frac{P_{s,h}}{P_s \times P_h},
\end{equation}
where $P_s$ and $P_h$ are the probability of detecting a signal or herald photon from a single laser pulse, and $P_{s,h}$ is the probability of detecting both at the same time. For completely uncorrelated light, $P_{s,h} = P_s \times P_h$, therefore $g^{(2)}_{s,h}=1$. For perfectly correlated light, $P_{s,h} = P_s = P_h$ returning $g^{(2)}_{s,h}=1/P_s$, so the $g^{(2)}_{s,h}$ can be arbitrarily high in the limit of low photon probability. For classical states of light, the $g^{(2)}_{s,h}$ is always between 1 and 2, a $g^{(2)}_{s,h}$ below 1 or above 2 is a sign of non-classical statistics. In figure~\ref{fig:countrates}(c) $g^{(2)}_{s,h}$ is measured as a function of laser pump power. The $g^{(2)}_{s,h}$ is over 2 for all pump powers, and exceeds 1000 in the low-power limit. The source therefore generates highly non-classical photon statistics which can be to distinguish the target from the background.

\subsection{Classical illumination}
To compare QI and CI, a classical light source with the same intensity, spectrum, polarization, and temporal profile as the signal photons is required. For simplicity, the signal photons themselves are used as the classical source, but instead of counting coincidences with the herald, single detection events are used. Due to imperfect heralding efficiency, the singles counts are around 13 times higher than the coincidence counts. Despite this, there is still a significant improvement in SNR when measuring conincidences compared to singles.

In the literature, classical illumination is often measured by splitting a thermal state ({\em e.g.} one mode of an \SPDC source) on a beam splitter and measuring correlations~\cite{Lopaeva2013}. While this would give an improved SNR, it comes at the expense of a reduced count rate. 
We therefore believe that counting single photons provides a more stringent comparison between QI and CI.

\subsection{Illumination apparatus}

\begin{figure}[h]
\center{\includegraphics[width=1.0\linewidth]{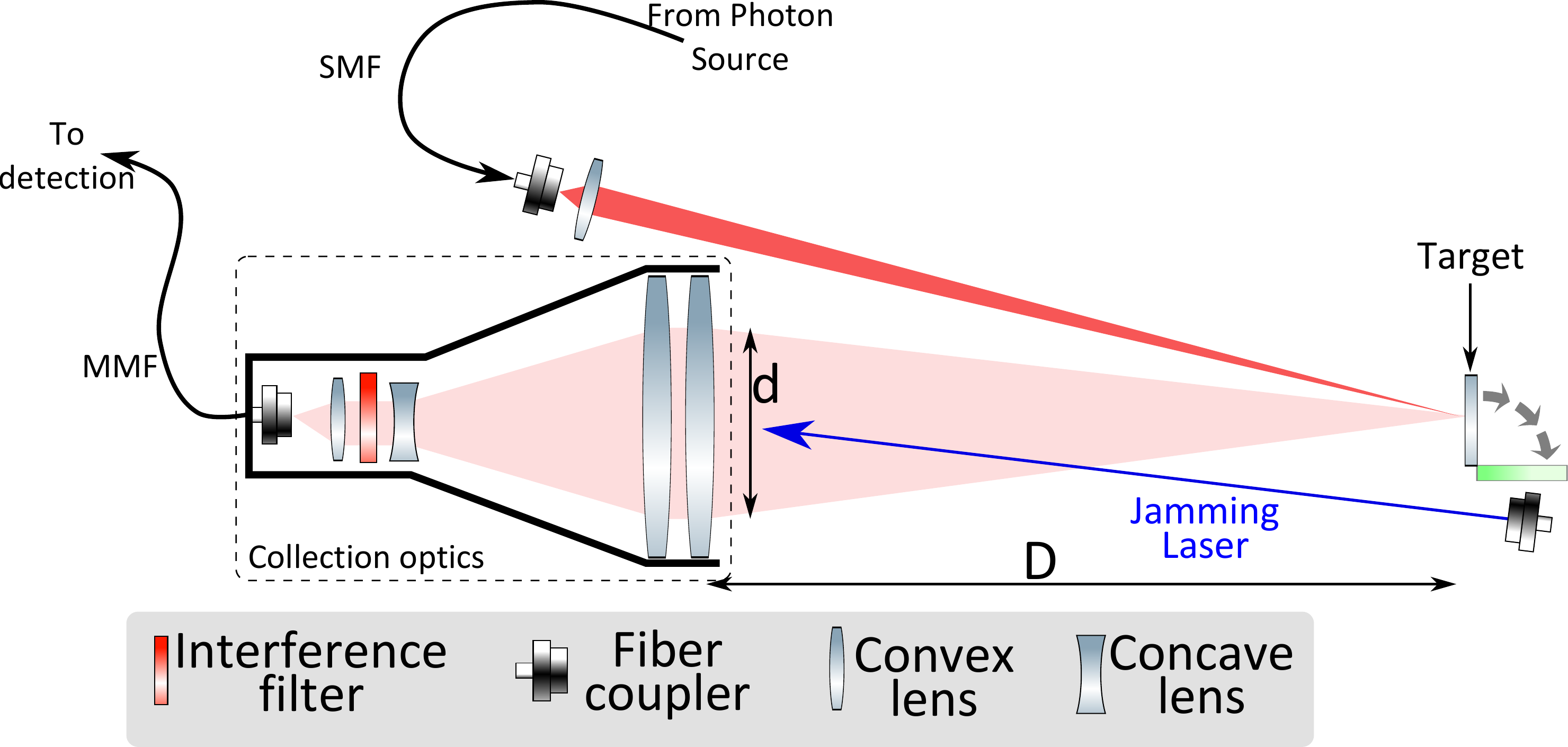}}
\caption{{\bf (a)} Schematic diagram of the illumination setup. The target is irradiated with photons from the source delivered via a single mode fiber (SMF) and scattered photons are imaged  by the collection optics onto the tip of a multimode fiber (MMF). The MMF delivers photons to the APD for detection. The collection optics are a distance $D$ form the target and the diameter of the mode that is collected is $d$. The ratio $d/D$ will determine the fraction of the scattered photons that can be collected (see text). An auxiliary laser beam (the jamming laser) is used to provide a background.}
\label{fig:setup_illumination}
\end{figure}

The apparatus used to illuminate and detect the target is shown in figure~\ref{fig:setup_illumination}. Light from the photon source is delivered to the apparatus via a single mode optical fiber (SMF) and is focused onto the target by an adjustable collimating lens. The target can be moved in and out of the beam using the shutter to test target recognition. The collection optics consist of a series of lenses designed to image the spot on the target onto the tip of a multimode optical fiber (MMF) which delivers the collected photons to an APD for detection. All of the collection optics are in a black metal tube, covered with black cloth to prevent room lights from entering the apparatus, and the MMF is sheathed in stainless steel and black rubber to prevent scattered photons entering the fiber through the cladding. Additionally, a narrowband optical filter (Semrock LL01-671) was placed in the collection optics to ensure that only light of the appropriate wavelength was collected.

The collection optics are placed a distance $D$ away from the target, and in this case $D=32$\,cm. The maximum diameter of a beam that can be focused onto the fiber is $d$: this can be measured experimentally by back-propagating a bright laser beam through the MMF and measuring the resulting beam-waist returning $d\simeq3$\,cm. A ratio of solid angles is used to estimate the maximum fraction $R_{max}$ of the scattered photons it would be possible to collect into the MMF:
\begin{equation}
\label{eq:collect}
R_{max} = \frac{\pi(d/2)^2}{2\pi D^2} = \frac{d^2}{8D^2}.
\end{equation}
Inserting experimental parameters returns $R_{max} = 1.1\times10^{-3}$. In making this simple estimate, it is assumed that light is equally scattered in all directions and that none is absorbed by the card. The real collection efficiency of the apparatus is measured by replacing the photon source with a bright beam and measuring the fraction of the power collected by the MMF returning a ratio of $R \simeq 3\times10^{-4}$, so around 1/3 of the available light is collected. 

\subsection{Jamming Laser}

To provide a controllable source of background illumination, a second laser referred to as the {\em jamming laser} is introduced. The jamming laser is a tunable pulsed laser (optical parametric oscillator) that is sent directly into the aperture of the collection optics. The wavelength and pulse arrival time are adjusted such that light from the jamming laser is spectrally and temporally indistinguishable from the photons arriving from the source. The power of the jamming laser is adjusted by a $\lambda/2$ plate and a polarizer (not shown in figure~\ref{fig:setup_illumination})

\section{Theoretical analysis}\label{sec:Theory}
In this section, a brief theoretical introduction is provided clarifying the key differences between QI and CI. We begin by defining the signal-to-noise ratio (SNR): The SNR is the ratio of photon detection events that occur due to photons from the illumination beam to the number of detection events due to the background alone. In practice, measured count rates with the target in place include both signal and background, whereas with the target removed they contain only background. An alternative definition of SNR is therefore given by the equation:
\begin{equation}\label{eq:SNR}
SNR = \frac{N_{in} - N_{out}}{N_{out}},
\end{equation}
where $N_{in}$ and $N_{out}$ are the number of detection events with the target in/out. Here the definition of a ``detection event'' is deliberately vague because for QI a detection event refers to a coincident detection between the signal and herald detectors, and for CI it refers only to single detections on the signal detector.

 The expected SNR values for both the CI and QI can then be calculated, in each case, we will consider an individual time bin, and discuss the {\em probability} of detecting a photon in that bin. The temporal duration of the time-bin and the number of bins per second will depend upon experimental conditions. Here, a bin width of 2\,ns is chosen because of temporal jitter of the detectors and there are $80\times10^6$ bins per second due to the repetition rate of the laser.
 
 \subsection*{Classical SNR}
 
In the absence of background, the probability of detecting a signal photon in a given time-bin is $\eta \times P_s$. Where, $P_s$ is the probability that a photon is generated in the source and $\eta$ is the overall collection efficiency which incorporates all collection losses and the detector efficiency. In the presence of background, the probability of detecting a background photon in the time-bin is $P_b$. The signal to noise ratio is therefore simply given by:
\begin{equation}
SNR_c = \frac{\eta P_s}{P_b}.
\end{equation}
 
 \subsection*{Quantum SNR}
The quantum SNR is similar. In the absence of background, the probability of detecting a signal-herald coincidence is $\eta P_{s,h}$. As before, a background photon is detected with probability $P_b$. The probability of accidentally detecting the background photon in coincidence with a herald photon is given by $P_{b,h} = P_hP_b$, where $P_h$ is the probability of detecting a herald photon. The SNR is therefore given by:
\begin{equation}\label{eg:SNRq}
SNR_q = \frac{\eta P_{s,h}}{P_hP_b}.
\end{equation}

 \subsection*{Quantum Enhancement}
To quantify the advantage of QI over CI, the ratio of the two SNRs is represented as the {\em Quantum Enhancement Factor (QEF)}:
\begin{equation}\label{eq:QEF}
{\mathrm{QEF}} = \frac{SNR_q}{SNRc} = \frac{\frac{\eta P_{s,h}}{P_hP_b}}{\frac{\eta P_{s}}{P_b}} =  \frac{P_{s,h}}{P_sP_h} = g^{(2)}_{s,h}.
\end{equation}

This yields the interesting result that the enhancement of QI over CI is directly related to the $g^{(2)}_{s,h}$ of the photon source. This will be shown experimentally in section~\ref{sec:QEF_Results}. We note that this analysis yields similar results to those shown in reference~\cite{Lopaeva2013} where the spatially-resolved multi-pixel quantum sensor in \cite{Lopaeva2013} has been replaced by a single temporally-resolved detector in this case. Two other important results derive from this simple model: both $\eta$ and $P_b$ drop out of the equation as they are present in both the classical and quantum SNRs. This shows that, regardless of increasing loss and background, QI will always have an SNR improvement over a classical source of the same intensity.

\section{Results}

\subsection{Comparison of QI and CI in isolated environment}

As an initial test of the \QI protocol, the entire illumination apparatus is placed inside a dark box, and the jamming laser is switched off. The pump power to the source is adjusted and photons are counted for 30 seconds with/without the target in place. To implement the \QI protocol, photons are counted in coincidence with the herald photon, and to implement \CI the photons are simply counted and correlations with the herald are ignored. In this way QI and CI can be compared side-by-side in identical conditions. 

\begin{figure}[h]
\center{\includegraphics[width=0.85\linewidth]{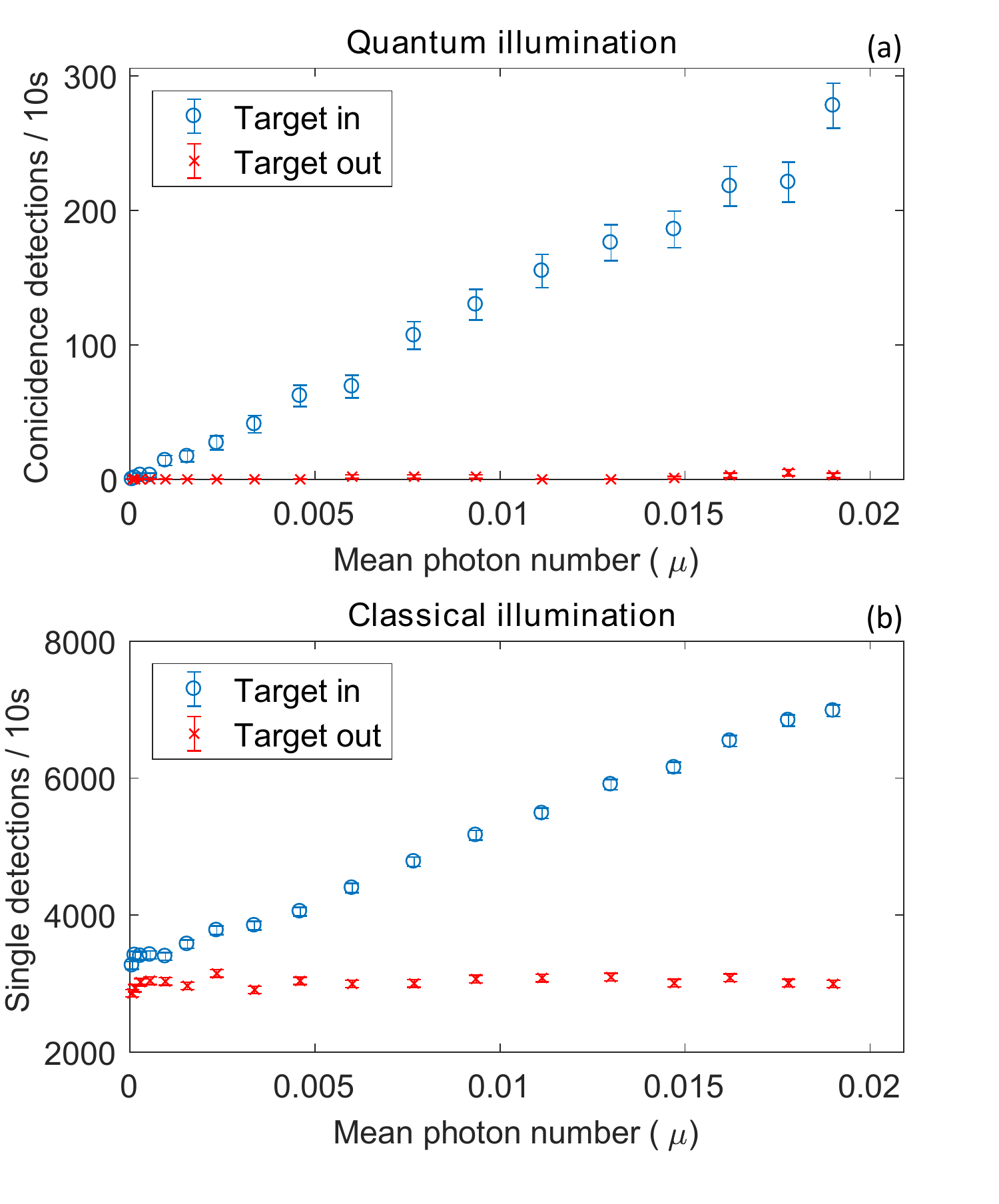}}
\caption{Target illumination in an isolated environment {\em i.e.} with minimal background light. {\bf (a)} QI --- Coincidence counts per 10\,s as a function of photon flux. {\bf (b)} CI --- singles counts per 10\,s as a function of photon flux. QI shows a clear benefit over CI as coincidences can be used to minimize detector noise. SNR should be interpreted as $(blue - red)/red)$.}
\label{fig:Background_Comparison}
\end{figure}

In figure~\ref{fig:Background_Comparison}, CI and QI are compared in the case of very low background light. In figure~\ref{fig:Background_Comparison}(a) coincidence counts are plotted with the target in/out as a function of the mean photon number: as expected a linear increase in coincidence counts is shown with the target in place. With the target out, almost no coincidence counts are measured, QI therefore provides an excellent signal-to-noise ratio. By contrast, in figure~\ref{fig:Background_Comparison}(b) it can be seen that, even with the target out, a significant number of single detections occur, predominately due to electrical noise in the detector. This results in a reduced SNR when compared to QI. Nevertheless, the classical SNR is easily high enough to distinguish the presence of the target so there is no compelling case for the use of QI in the limit of low background light. 

\subsection{Comparison of QI and CI in the presence of a strong background}
\begin{figure}[h]
\center{\includegraphics[width=0.85\linewidth]{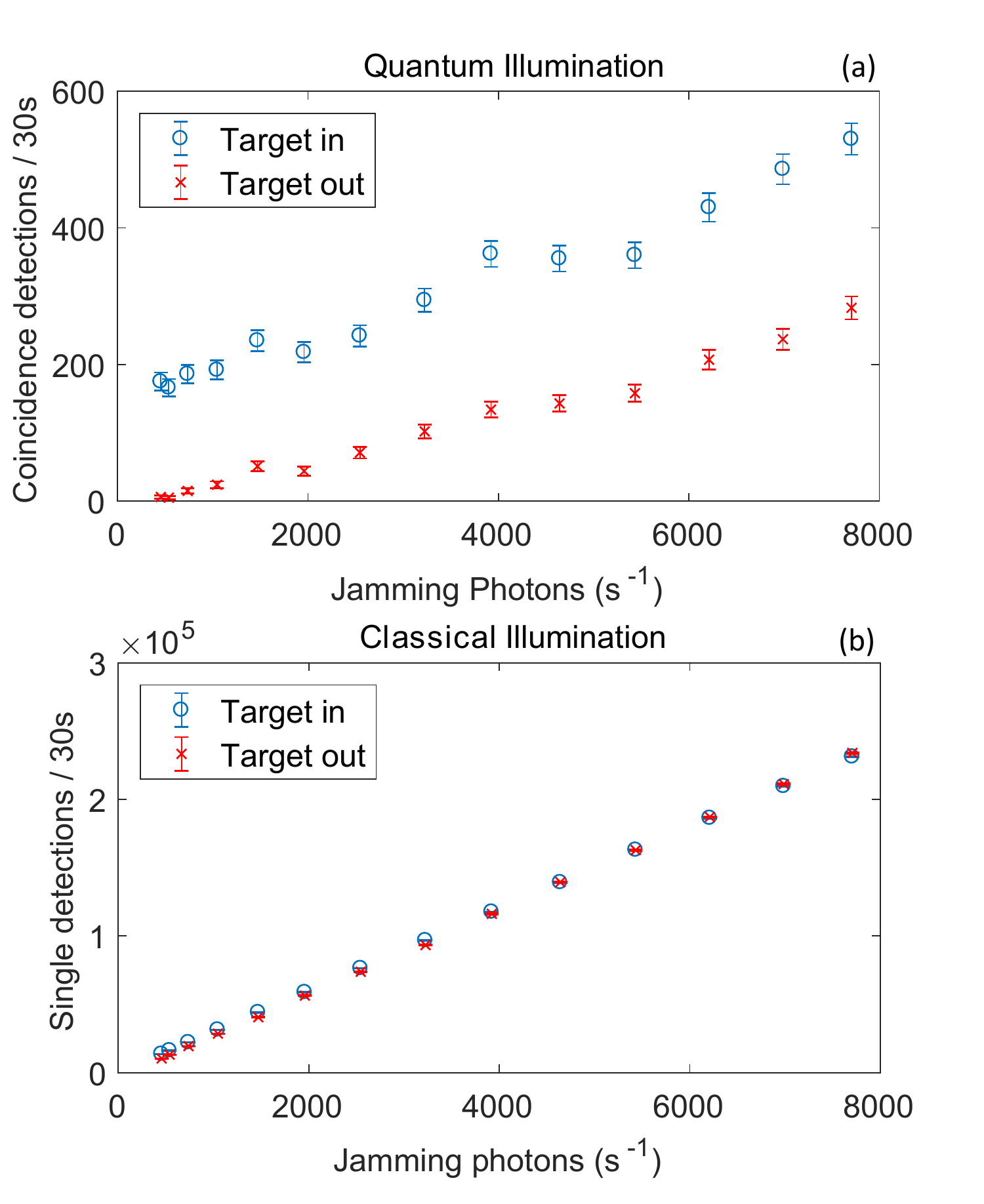}}
\caption{Target illumination in the presence of the jamming laser.  {\bf (a)} QI --- Coincidence counts per 30\,s as a function of jamming photon flux. {\bf (b)} CI --- singles counts per 30\,s as a function of jamming photon flux. QI shows a clear benefit over CI in terms of SNR. Mean photon number transmitted to the target $\mu \simeq 5.8\times10^{-3}$.}
\label{fig:jamming_Comparison}
\end{figure}

The study of QI is continued by introducing the {\em jamming laser} --- an auxiliary laser beam which is shone directly into the collection optics. This effectively simulates a target which is hidden in a strong background. The intensity of the jamming laser is adjusted using the motorized $\lambda/2$ plate, singles and coincidence counts as a function of jamming laser intensity are plotted in figure~\ref{fig:jamming_Comparison}. This provides a compelling case for QI: a clear increase in coincidence counts is evident with the target in [\ref{fig:jamming_Comparison}(a)] whereas there is no observable difference in the singles counts [\ref{fig:jamming_Comparison}(b)]. In the limit of a strong background, both QI and CI are sensitive to jamming, but QI is far less sensitive due to quantum correlations. This enables a clear signal from QI which could not be observed with classical light.

\subsection{Quantifying the quantum advantage}\label{sec:QEF_Results}
\begin{figure}[h]
\center{\includegraphics[width=0.8\linewidth]{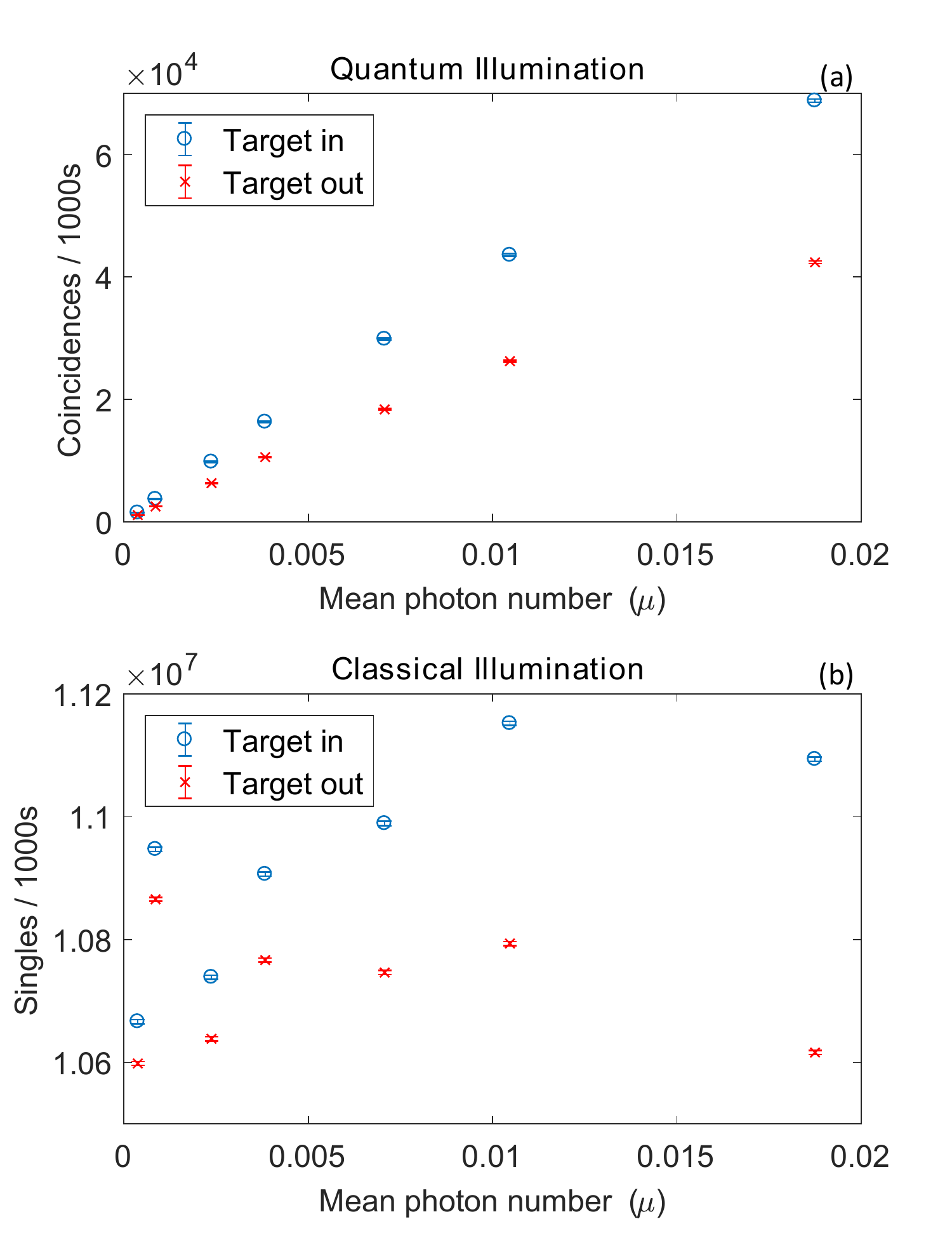}}
\caption{Target illumination in the presence of the jamming laser ($\sim10,000$\,s$^{-1}$) with (blue circles) and without (red crosses) the target in place. {\bf (a)} QI --- Coincidence counts per 1000\,s as a function of mean photon number generated by the source. {\bf (b)} CI --- singles counts per 1000\,s as a function of mean photon number. Note that the CI data is much more scattered than the QI despite both being acquired simultaneously. This is due to fluctuations in the jamming laser intensity. Because the SNR is far lower in CI, this data is more susceptible to these fluctuations.  }
\label{fig:CIQISNR}
\end{figure}

\begin{figure}[h]
\center{\includegraphics[width=0.85\linewidth]{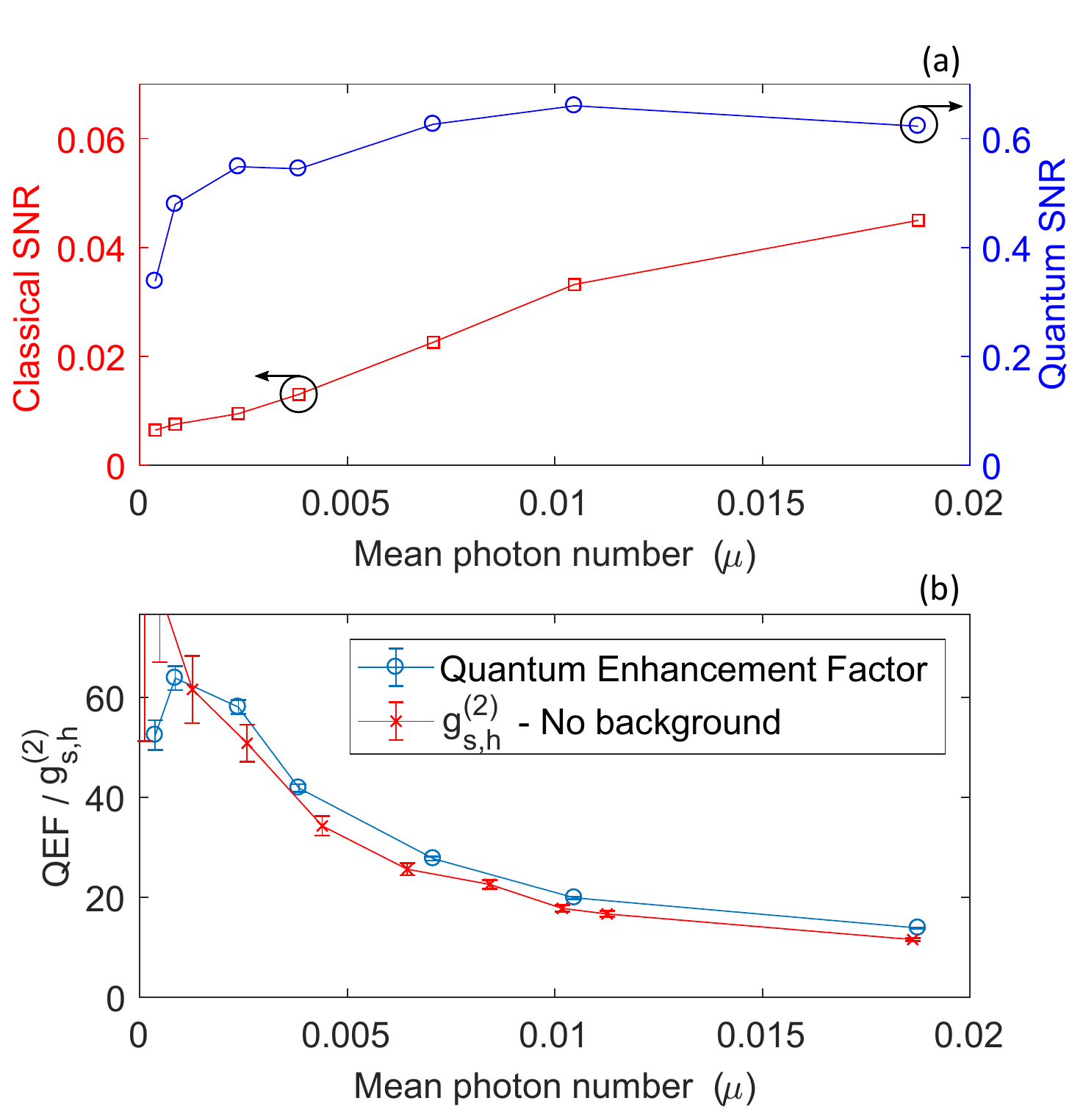}}
\caption{{\bf (a)} Signal-to-noise ratio for CI (red squares) and QI (blue circles) as a function of mean photon number in the presence of the jamming laser ($\sim10,000$\,s$^{-1}$). Note the order-of-magnitude difference in scale. {\bf (b)} The ``Quantum Enhancement Factor'' --- the ratio of quantum SNR to classical SNR --- plotted against mean photon number (blue circles). The $g^{(2)}_{s,h}$ of the photon source, after the signal photon has been scattered from the target, has been plotted for comparison (red crosses). }
\label{fig:QEF}
\end{figure}

In figure~\ref{fig:CIQISNR}, singles and coincidences are measured with the target in and out as a function of the mean photon number generated by the source. The jamming laser intensity is fixed such that $\sim10,000$ photons per second are detected. The photon counts are monitored over a 2000\,s period at each mean photon number; in order to mitigate long-term drifts, the target shutter is toggled every second such that  1\,s of data is taken with the target in followed by 1\,s with the target out. The SNR is then calculated according to equation~\ref{eq:SNR}, which is plotted, for both QI and CI, in figure~\ref{fig:QEF}(a).

In figure~\ref{fig:QEF}(a), it can be seen that the classical SNR increases approximately linearly as the mean photon number is increased; this is to be expected because as more photons are sent, more are received by the detector. By contrast, after an initial increase, the quantum SNR levels off and becomes independent of mean photon number. This is because, as $\mu$ increases, not only are more signal photons generated, but also more herald photons so the number of accidental coincidences between herald and background increases in proportion to the signal-herald coincidences. This can be seen in equation~\ref{eg:SNRq}. The ``Quantum Enhancement Factor'' (QEF) is then calculated as the ratio of quantum to classical SNR, and is plotted in figure~\ref{fig:QEF}(b). As the quantum SNR is broadly independent of $\mu$ and the classical SNR is proportional to $\mu$ the QEF is largest at low mean photon numbers, as is expected from equation~\ref{eq:QEF}.

Equation~\ref{eq:QEF} predicts, using a simple model, that the QEF would be exactly the $g^{(2)}_{s,h}$ of the photon source. Here the $g^{(2)}_{s,h}$ is measured as a function of $\mu$ by blocking the jamming laser and measuring singles and coincidences for 60\,s at each value of $\mu$. The results, shown in red in figure~\ref{fig:QEF} are in excellent agreement with the QEF data verifying the simple model. It should be noted here that the  $g^{(2)}_{s,h}$ only reaches around 60 compared to $\sim1000$ in figure~\ref{fig:countrates}(c), this is because the  $g^{(2)}_{s,h}$ is measured after the photon has been scattered from the target introducing a large degree of loss compared to the direct measurement made in figure~\ref{fig:countrates}(c). In this case stray light and detector dark counts become significant and degrade the $g^{(2)}_{s,h}$. The $g^{(2)}_{s,h}$ measured in figure~\ref{fig:countrates}(c) can be considered an upper-bound of the potential QEF in the limit of perfect detectors.

\subsection{Imaging}

\begin{figure}[h]
\center{\includegraphics[width=0.9\linewidth]{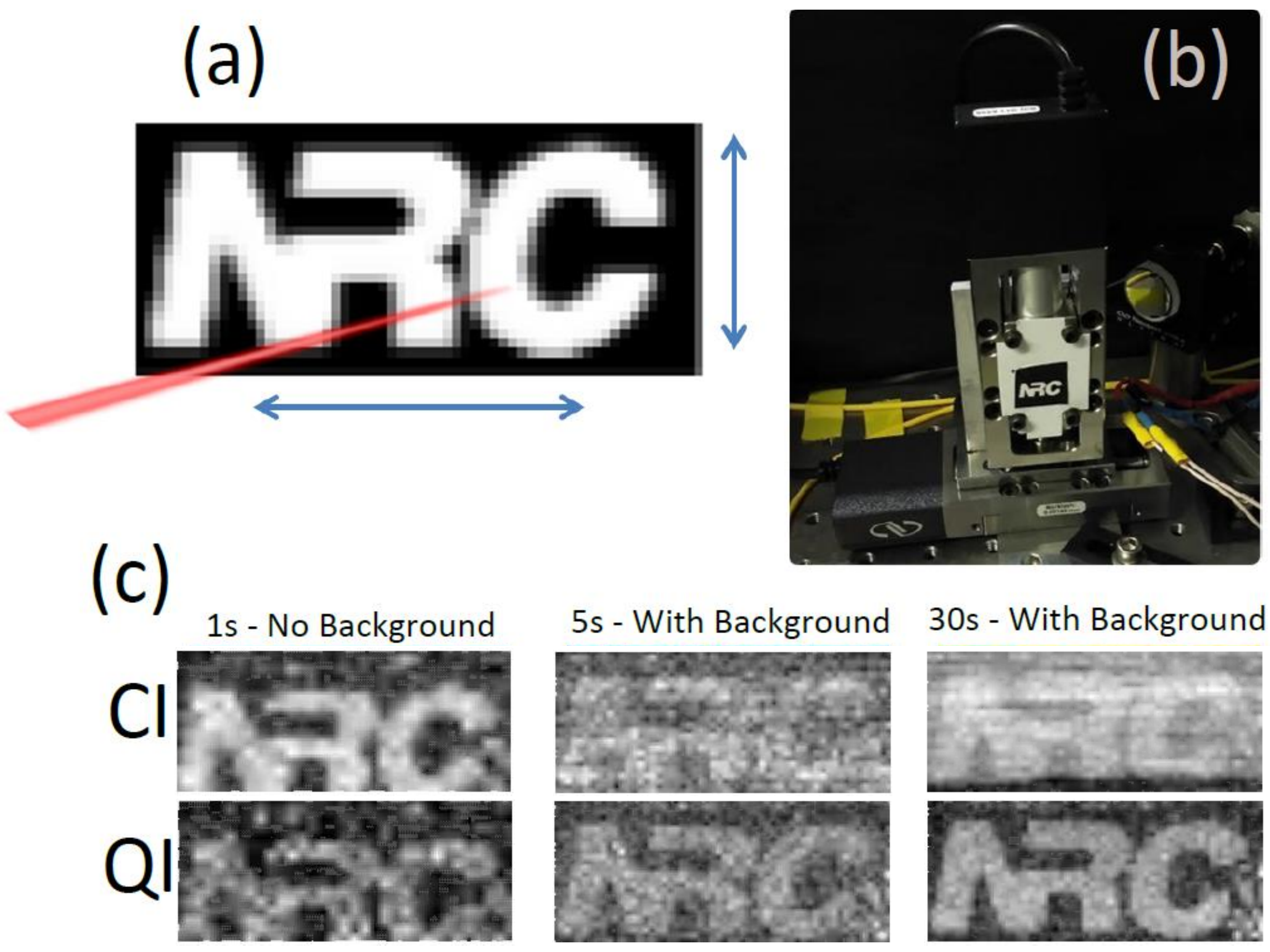}}
\caption{Imaging apparatus and results. {\bf (a)} Conceptual setup --- The photon beam remains fixed while the target is raster-scanned to build up an image. {\bf (b)} Photograph of the apparatus. A pair of motion stages are used to move the target up/down and left/right. {\bf (c)} Typical images taken using CI (top) and QI (bottom). The first pair of images are taken in the absence of background; all others have a background detection rate of $\sim 14,000\,$s$^{-1}$. The integration time per pixel is indicated above each image pair. Note: a 4-point interpolation is used to smooth the pixelated image. Mean photon number transmitted to the target $\mu \simeq 7.9\times10^{-3}$. }
\label{fig:Imaging}
\end{figure}

In order to gain a visual appreciation of the advantage of QI, some simple imaging measurements are shown. It should be noted that these are not single-shot images because a multi-pixel quantum detector is not used. Instead the laser beam remains fixed and the target is moved; by raster-scanning an image is built up. In this case, the target is a white NRC logo on a black background. A schematic and photograph of the raster-scanning setup is shown in figure~\ref{fig:Imaging}(a) and (b). 

Typical images obtained using the raster-scanning technique are shown in figure~\ref{fig:Imaging}(c). Initially the jamming laser is switched off, and photon counts are integrated for 1\,s at each pixel. In this configuration, CI produces a sharper image than QI. However, when the jamming laser is added with intensity set such that $\sim14,000$ background photons per second are detected, the QI image is sharper than the CI. This confirms an observation made previously: a strong background is required in order for QI to have a significant benefit over CI. Further images are taken with 2, 5, 7, 10, 20 and 30 second integration times per pixel. In figure~\ref{fig:Imaging}(c) we show the 5\,s and 30\,s images to draw an interesting comparison. Note that the sharpness of the 30\,s CI image is comparable to that of the 5\,s QI image, so similar images require less time using QI compared to CI. These results are not a surprise given previous quantitative measurements, but they provide a qualitative visual illustration of the benefits of QI.

\section{Discussion and Outlook}
This demonstration was designed to be a proof-of-principle setup in a controlled laboratory environment with only 32\,cm between the target and the detector. Nevertheless it has helped to identify regimes in which QI may have a significant benefit over CI. It is important to note that, at all points, QI has been compared with a CI source {\em of the same intensity}; greater sensitivity in CI can always be achieved by increasing the intensity of the source. Also, somewhat counter-intuitively, a substantial background is required in order to reveal the advantages of QI. Therefore QI will primarily be of interest for applications where a target must be detected in a high background while sending as few photons as possible. 

An important result of this work is the direct relationship between the  degree of second order coherence of the pair source ($g^{(2)}_{s,h}$) and the quantum enhancement factor (QEF), which enumerates the advantage of using the pair source instead of an attenuated classical beam of the same intensity. Since the $g^{(2)}_{s,h}$ of the source is inversely proportional to the probability of generating a photon pair, the QEF is highest when the mean number of photons per mode $\mu$ is lowest. Conversely, of course, the absolute SNR of both CI and QI will increase as more photons illuminate the target. So, to achieve high SNR and high QEF, one must increase the photon production rate while also increasing the number of modes populated so to keep $\mu$ low. In practical terms, this will involve the use of a continuous wave pump laser and high-speed single photon detectors to increase the number of temporal modes as well as leveraging other degrees of freedom such as polarization or frequency.  

\section{Acknowledgments}
This work was supported by the Defence R\&D Canada's Air Reach Data Fusion project. The authors are grateful to Jennifer Erskine, Khabat Heshami, Philip Bustard, Denis Guay, Doug Moffatt, Adrian Pegoraro, Paul Hockett, and Rune Lausten for support and insightful discussions.



\bibliography{QI_Papers}{}
\end{document}